# Tunable ultrabroadband hybrid THz emitter combining a spintronic THz source and a GaSe crystal

Afnan Alostaz[1], Oliver Gueckstock[1], Jungwei Tong[1], Jana Kredl[2], Chihun In[1], Markus Münzenberg[2], Tom S. Seifert*[1]

[1]Freie Universität Berlin, Germany

[2]Universität Greifswald, Germany

*Email: tom.seifert@fu-berlin.de

**Abstract**

Linear terahertz time-domain spectroscopy (THz-TDS) is a sensitive probe for material characterization including thickness measurements of thin layers. These applications critically rely on a sufficiently large bandwidth, which is not straightforwardly available in typical THz-TDS systems. Here, we introduce a hybrid THz-emitter concept based on a spintronic THz emitter that is deposited onto a thin free-standing GaSe nonlinear crystal. By tuning the parameters of this hybrid emitter, we generate an ultrabroadband spectrum covering the full range from 1 to 40 THz without any gaps at high spectral amplitudes, resulting in ultrashort THz-pulse durations of only 32 fs. Finally, we demonstrate the straightforward tunability of the carrier-envelope phase from unipolar or bipolar THz pulses with ultrashort duration.

## 1. Introduction

Terahertz (THz) radiation is a sensitive probe of many low-energy excitations in solids, liquids and gases [1]. Prominent examples in condensed matter include lattice vibrations, spin waves and intraband-type transport. This sensitivity to low-energy excitations enables broad applications of linear THz spectroscopy in, for example, material identification by THz refractive-index determination [2] and thickness measurements of thin layers such as polymers [3] or paints [4].

Importantly, for those tasks, THz time-domain spectroscopy (THz-TDS) is ideally suited as it allows one to record signal amplitude and phase simultaneously [2]. For THz generation, THz-TDS systems driven by nanojoule femtosecond laser pulses often use optical rectification in nonlinear crystals such as GaP, ZnTe and GaSe, or photoconductive switches [1, 5]. Typically, these THz sources work in a relatively narrow frequency range that is limited by phase-matching conditions or phonon absorption bands [6]. Accordingly, linear THz spectroscopic measurements are common in the range 1-5 THz for ZnTe and GaP crystals as well as for photoconductive emitters [2], or at 10-40 THz for GaSe crystals [7].

Recently, ultrabroadband spintronic THz emitters (STEs) based on thin metallic multilayers emerged as a promising THz-emitter platform that, in particular, emit in the range 1-15 THz [8, 9]. STEs rely on ultrafast laser-driven out-of-plane spin currents $j_s$ that get converted inside the paramagnetic metal layers P into in-plane charge currents $j_c$. The latter radiate electromagnetic waves with THz frequencies into the far-field. The STE's insensitivity to the driving wavelength and the ability to generate ultrabroadband THz pulses covering the entire THz range, lead to detailed insights into the THz refractive index of a polymer [9] or into transverse THz magnetotransport in magnetic metals [7] at 1-40 THz.

However, while STEs can excellently cover the interval 1-15 THz, which includes the THz gap at 5-10 THz, the spectral amplitude for frequencies above 15 THz decreases rapidly. A boost of these high-frequency components could strongly reduce acquisition times in linear THz-TDS measurements and produce ultrashort THz pulses that are well suited as probe pulses in experiments as diverse as transient THz conductivity measurements [10], THz-pump photoelectron-probe experiments [11] or THz-driven scanning-tunneling microscopy [12].

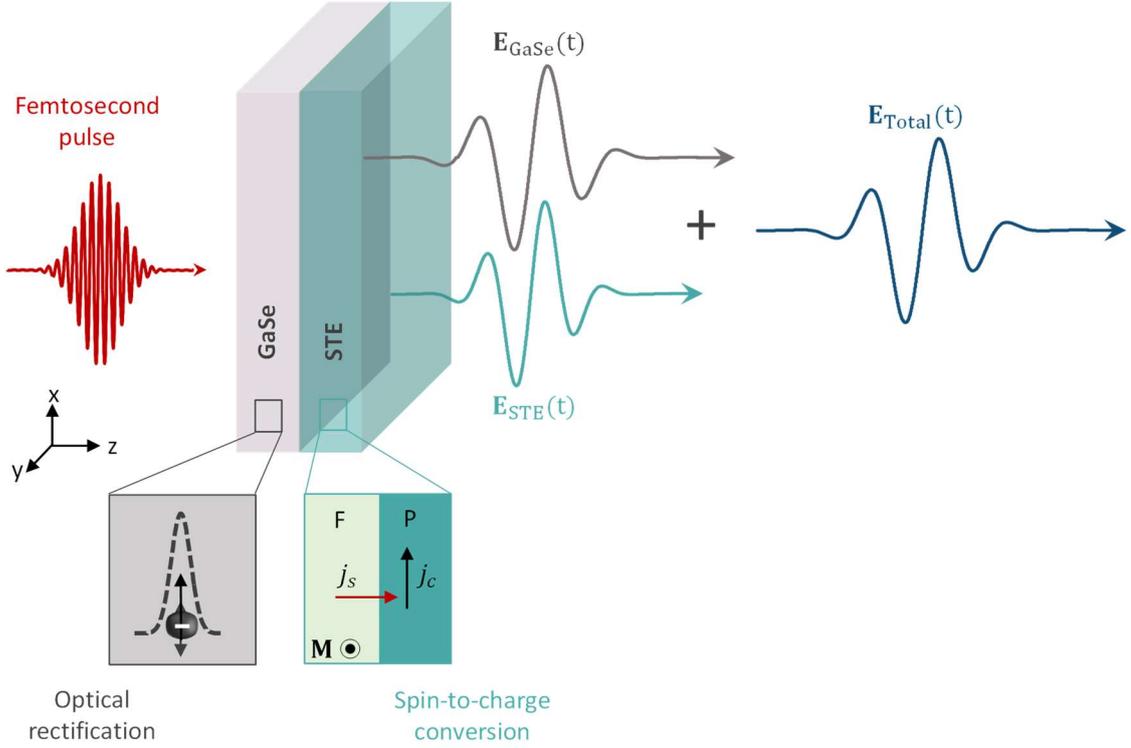

**Figure 1: Hybrid THz emitter concept.** The emitter consists of a thin GaSe crystal with a spintronic THz emitter (STE) grown on top. Upon excitation by a femtosecond near-infrared laser pulse, optical rectification inside the GaSe crystal generates a THz pulse with electric field $\mathbf{E}_{\mathrm{GaSe}}(t)$ (grey). The STE thin-film stack consists of a metallic ferromagnetic layer F and two layers of paramagnetic metals P. After optical excitation, a spin current $j_s$ is injected into the adjacent F. Spin-orbit interaction converts $j_s$ into an in-plane charge current $j_c$ that leads to the emission of a THz pulse with field $\mathbf{E}_{\mathrm{STE}}(t)$ (green). Both THz electric fields add up to the total emitted THz pulse with electric field $\mathbf{E}_{\mathrm{Total}}(t)$.

Here, we present a hybrid THz emitter based on a thin free-standing GaSe crystal, on top of which a STE is deposited (see Fig. 1). The basic idea behind the hybrid emitter relies on superimposing the generated THz pulses from the STE and the GaSe. Importantly, the phase-matching conditions and the angular dependence of the nonlinear coefficient in GaSe allow for tuning its emission with the angle of incidence of the pump, the sample azimuth and the pump polarization [13]. Along with the polarity of the STE pulse, which can be set by an external magnetic field, one can tune the resulting superimposed THz pulse in duration and shape.

In this way, we generate THz pulses that cover the full range 1-40 THz with high spectral amplitudes, achieve shortest THz pulse durations of only 32 fs and tune the carrier-envelope phase to yield bipolar or unipolar ultrashort THz pulses by simply reversing the external magnetic field.

## 2. Experiment design

Our hybrid emitter consists out of a large-area and free-standing GaSe crystal (z-cut) with a thickness of 30 µm, on top of which a spintronic THz emitter with stacking order W(2 nm)|$Co_{40}Fe_{40}B_{20}$ (1.8 nm)|Pt(2 nm) is deposited (see Methods). The in-plane sample magnetization **M** is saturated by a static external magnetic field of about 40 mT (Fig. 1).

In the experiment, the near-infrared pump pulse (duration 10 fs, central wavelength 800 nm, repetition rate 80 MHz, pulse energy 5 nJ) is incident from the GaSe side. The sample can be rotated about its azimuth $\varphi$ and about an in-plane axis (p direction) to vary the angle of incidence $\theta$ of the pump. The sample orientation corresponding to $\varphi = 0°$ is chosen by setting GaSe to its maximum THz emission under normal incidence (see Fig. 1).

To detect the generated p-polarized THz pulses $S^\pm(t)$, we use electro-optic sampling [14] in a 250 µm-thick GaP(110)-crystal with probe pulses (1 nJ energy) from the same laser. The STE and GaSe THz signals are obtained from the THz-signal components odd and even in **M**, i.e., $(S^+ \mp S^-)/2$ respectively. All experiments are performed under a dry-air atmosphere. Further details about the experimental setup can be found in the Methods and in Ref. [9].

To extract the THz electric field $E(t)$ at the detector position, we use an inversion procedure that is based on deconvolution of the response function of the GaP detection crystal $R(t)$ from the detected THz signals $S(t)$, i.e, by solving $S(t) = R(t) * E(t)$ for $E(t)$ [15, 16].

To determine the duration of the THz electric field, we first calculate the intensity envelope $|A(t)|^2$, where the complex-valued $A(t) = (\mathcal{H}E)(t)$ is given by the Hilbert transformation $\mathcal{H}$ of $E(t)$. Second, we calculate the temporal width $\langle \Delta t \rangle$ of $|A(t)|^2$ as the standard deviation by [17]

$$\langle \Delta t \rangle = \sqrt{\langle t^2 \rangle - \langle t \rangle^2} \tag{1}$$

where

$$\langle t^n \rangle = \frac{\int_{-\infty}^{\infty} dt\, t^n\, |A(t)|^2}{\int_{-\infty}^{\infty} dt\, |A(t)|^2}. \tag{2}$$

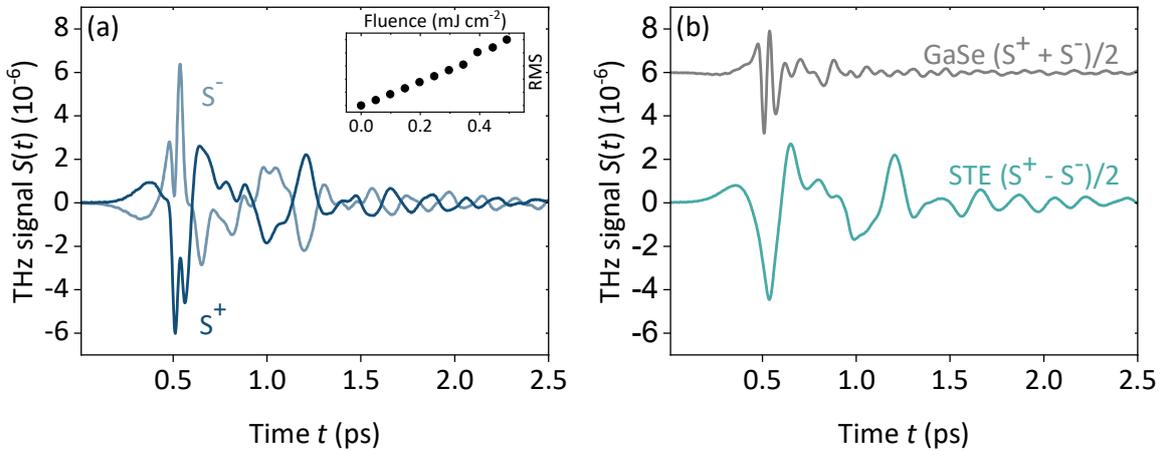

**Figure 2: Raw THz-emission signals from the hybrid emitter.** (a) THz-emission signal polarized perpendicular to the in-plane sample magnetization **M** from the hybrid emitter for magnetization directions $\pm$**M**. The inset shows the root-mean square (RMS) of the THz-emission signal vs the pump fluence. (b) Extracted STE and GaSe THz signals from the data in panel (a). The two signals are, respectively, the components odd and even in **M**, i.e., $(S^+ \mp S^-)/2$. The GaSe signal is vertically offset for clarity.

### 3. Results and discussion

Figure 2 shows raw THz-emission signals $S^\pm(t)$ from the hybrid emitter under normal incidence ($\theta = 0°$), with a linear pump polarization of 45° and at an azimuth of $\varphi = 0°$. Upon reversing the sample magnetization **M**, marked changes in the THz waveform appear. We extract the GaSe and STE contributions to the THz waveform by adding and subtracting the waveforms for $\pm$**M**, respectively (Fig. 2b). Indeed, two waveforms emerge from this procedure that closely resemble typical THz-emission signals from each of the two emitters [9, 13]. The inset in Fig. 2 confirms a linear fluence regime. These findings indicate the general feasibility of our hybrid-emitter concept.

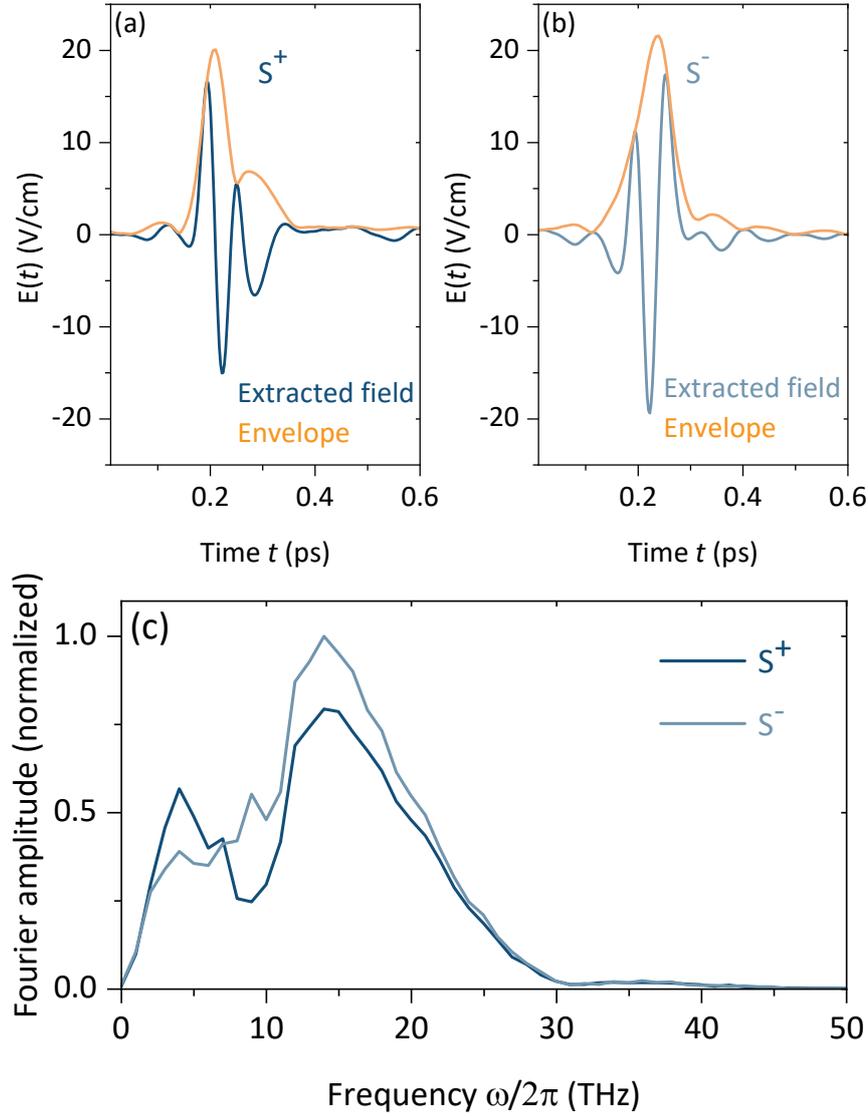

**Figure 3: Extracted THz-electric-field traces from the hybrid emitter.** (a) Terahertz electric field retrieved from the corresponding THz signals for $\theta = 0°$ and $\varphi = 0°$ for magnetization $+\mathbf{M}$ and (b) magnetization $-\mathbf{M}$. (c) THz spectra obtained by Fourier transformation of the corresponding field traces in panels (a) and (b). Both spectra are normalized by the same value.

In Figure 3, we show the extracted THz electric fields reaching peak values of about 20 V/cm in the detection focus. The retrieved amplitude envelopes (Fig. 3a and 3b) for opposite sample magnetizations demonstrate the tunability of the hybrid emitter by constructive or destructive superposition of the STE and GaSe THz pulses. This notion is further confirmed by the spectra (Fig. 3c) that already cover the entire range 1-40 THz even before optimization of the emitter geometry.

Based on these findings, we perform an optimization of the total THz emission from the hybrid emitter by variation of the angle of incidence $\theta$ and the sample azimuth $\varphi$. Remarkably, this procedure results in an ultrashort THz electric field with a duration of the intensity envelope of only 32 fs (Fig. 4a) with an ultrabroad spectrum (Fig. 4b). Thus, the hybrid emitter indeed unifies the advantages of the STE and GaSe within one device, where the STE covers the range 1-10 THz and GaSe covers 10-40 THz efficiently.

Eventually, we demonstrate tuning of the THz-pulse shape from unipolar to bipolar (Fig. 5) by reversing the direction of the STE magnetization from $+\mathbf{M}$ (Fig. 5a) to $-\mathbf{M}$ (Fig. 5b), while maintaining the ultrashort THz-pulse duration. This result paves the way towards ultrabroadband applications that are sensitive to the THz carrier-envelope phase such as THz-streaking of electron beams [18] or THz high-harmonic generation [19].

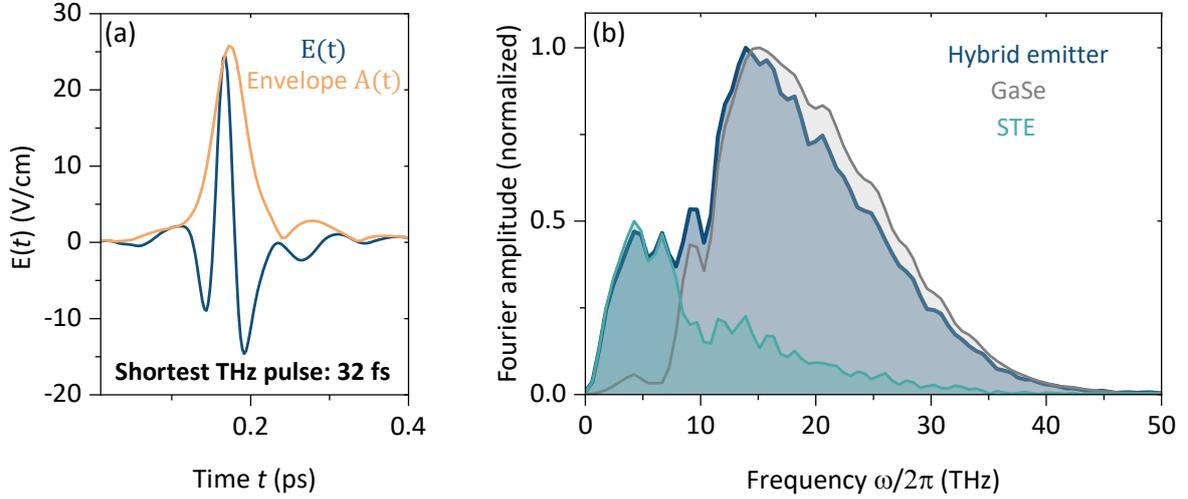

**Figure 4: Optimized THz-electric-field transient from the hybrid emitter.** (a) Extracted THz electric field of the shortest pulse yielding a pulse duration of 32 fs with respect to the intensity envelope $|A(t)|^2$ (see Eq. 2) at the detector position. To generate this shortest pulse, we set the hybrid THz emitter to $\theta = 40°$ and $\varphi = 60°$, and the pump polarization to 45°. (b) THz spectra of the pulse in panel (a) together with the respective contributions of the STE and the GaSe crystal.

We note that a future optimization of the heat dissipation in our hybrid emitter by, for instance, contacting it to a thicker substrate could further boost the THz amplitude. We expect an increase of about a factor of 10 in field from a comparison to STEs grown on a sapphire substrate [9].

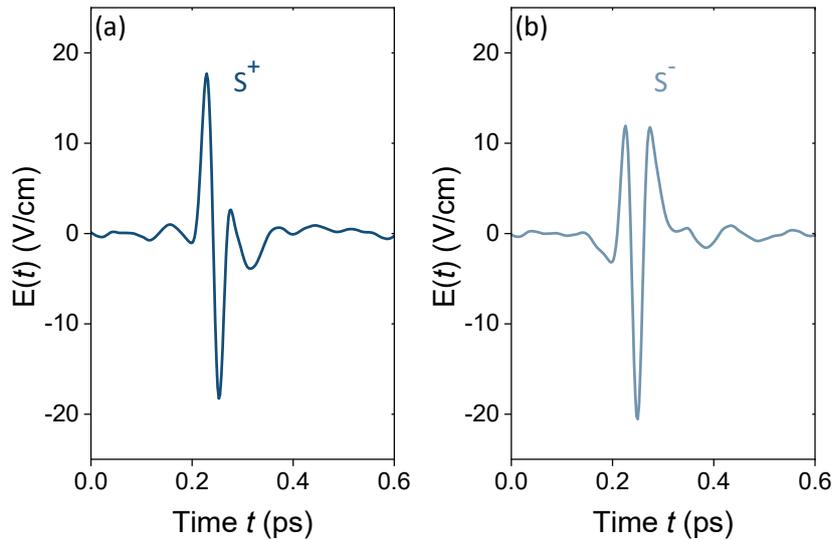

**Figure 5: Tuning from unipolar to bipolar THz pulses.** (a) Electric field of the THz pulse at the detector position for magnetization $+\mathbf{M}$ and (b) magnetization $-\mathbf{M}$. The magnetization direction permits tuning from a bipolar [panel (a)] to a unipolar [panel (b)] pulse shape. The emitter settings are $\theta = 40°$ and $\varphi = 30°$.

## 4. Conclusions and outlook

In summary, we demonstrate a hybrid-emitter concept that unifies the strengths of STEs and GaSe crystals as THz emitters in one device. Their combination results in an ultrabroadband THz source that can cover the range 1-40 THz at spectral amplitudes remaining above 3% of the spectral peak amplitude that is found at 15 THz. The resulting THz pulses feature a pulse duration of only 32 fs at $\theta = 40°$ and $\varphi = 60°$, which makes them prime candidates for experiments probing ultrafast dynamics of diverse types of THz resonances in all phases of matter.

**Methods**

*S1 Deposition of the spintronic THz emitter.*

The STE-layer stack W(2 nm)| Co$_{40}$Fe$_{40}$B$_{20}$ (1.8 nm)|Pt(2 nm) was deposited on top of the 30 µm-thick GaSe crystal in a multi-cluster preparation tool under ultrahigh vacuum conditions at a base pressure below 5·10$^{-10}$ mbar. The W and CoFeB films were magnetron sputtered in an Ar atmosphere at a pressure of 5·10$^{-3}$ mbar (target composition Co$_{40}$Fe$_{40}$B$_{20}$). The Pt layer was deposited in-situ by e-beam evaporation in the UHV cluster with a base pressure of about 10$^{-9}$ mbar at a deposition rate of 0.02 nm/s.

*S2 Experimental setup.*

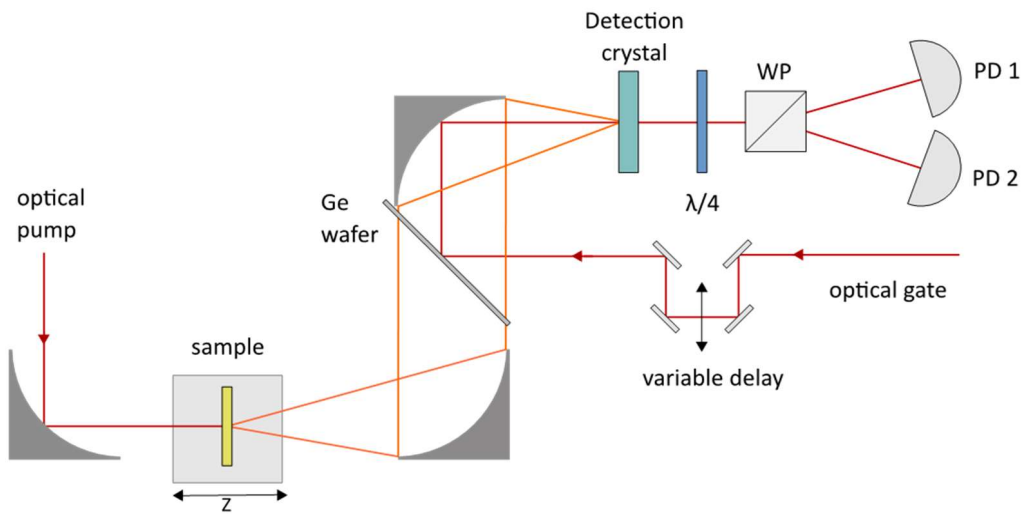

**Figure S1: Schematic of the THz spectrometer**. The sample is excited by an incident optical pump pulse and generates a THz pulse that propagates co-linearly with the optical gate pulse into a nonlinear detection crystal. Behind the detection crystal, the THz-induced probe ellipticity is recorded by a balanced detection scheme consisting of a quarter-wave plate (λ/4), a Wollaston prism (WP) and a pair of photodiodes (PD1/2). The figure is taken from [20] with kind permission of O. Gueckstock.